\documentstyle[12pt,epsf,epsfig,amssymb]{article}
\begin{document}                                                                                   
              
% JOURNAL MACRO :
% ===============
\def\Journal#1#2#3#4{{#1} {\bf #2} (#3) #4}
% JOURNAL NAMES :
% ===============
\def\NCA{\em Nuovo Cimento}
\def\NIM{\em Nucl. Instrum. Methods}
\def\NIMA{{\em Nucl. Instrum. Methods} A}
\def\NPB{{\em Nucl. Phys.} B}
\def\PLB{{\em Phys. Lett.}  B}
\def\PRL{\em Phys. Rev. Lett.}
\def\PRD{{\em Phys. Rev.} D}
\def\ZPC{{\em Z. Phys.} C}
\def\ZP{\em Z. Phys.}
\def\PR{\em Phys. Rev.}
\def\CPC{\em Comp. Phys. Commun.}
\def\SJNP{\em Sov. Journ. Nucl. Phys.}
\def\JETP{\em Sov. Phys. JETP}
\def\PREP{\em Phys. Rep.}
\def\ARNPS{\em Ann. Rev. Nucl. Part. Sci.}
\def\EPJC{{\em Eur. Phys. J.} C}
\def\NPBPS{{\em Nucl. Phys.} B (Proc. Suppl.)}
% D0 experiment :
% ===============
\def\D0{D\O}  \def\d0{D\O}

\title{\begin{flushright}{\large DESY--99--054}  \end{flushright}
\vspace*{2cm}
\LARGE \bf{
$W$ Production and the Search for Events with an Isolated High-Energy Lepton and
Missing Transverse Momentum at HERA}}

\date{}
\author{ZEUS Collaboration }
\maketitle
\begin{abstract}
\noindent
A search for the leptonic decays of $W$ bosons produced in the reaction \linebreak $e^{+}p\to e^{+}W^{\pm}X$ at a 
centre-of-mass energy of $300$~${\rm GeV}$ has been performed with the ZEUS detector at HERA using an integrated
luminosity of $47.7$~${\rm pb^{-1}}$. Three events consistent with $W\to e\nu$ decay are found, giving a cross 
section estimate of $0.9^{+1.0}_{-0.7}\pm0.2$~${\rm pb}$, in good agreement with the Standard Model prediction.
The corresponding $95\%$~C.L. upper limit on the cross section is $3.3$~${\rm pb}$. A search for
the decay $W\to \mu\nu$ has a smaller selection efficiency and yields no candidate events. Limits on the
cross section for $W$ production with large hadronic transverse momentum have been obtained.
A search for high-transverse-momentum 
isolated tracks in events with large missing transverse momentum yields results in good agreement with Standard Model 
expectations, in contrast to a recent report by the H1 collaboration of the observation of an excess of such events.
\end{abstract}
\pagestyle{plain}                   
\thispagestyle{empty}

% Author and institute lists :
% ============================
% Set page-numbering to Roman :
\clearpage
\pagenumbering{Roman}

\begin{center}                                                                                     
{                      \Large  The ZEUS Collaboration              }                               
\end{center}                                                                                       
  J.~Breitweg,                                                                                     
  S.~Chekanov,                                                                                     
  M.~Derrick,                                                                                      
  D.~Krakauer,                                                                                     
  S.~Magill,                                                                                       
  B.~Musgrave,                                                                                     
  A.~Pellegrino,                                                                                   
  J.~Repond,                                                                                       
  R.~Stanek,                                                                                       
  R.~Yoshida\\                                                                                     
 {\it Argonne National Laboratory, Argonne, IL, USA}~$^{p}$
% ...
\newpage                                                                                           
% ...
$^{  36}$ partially supported by the Foundation for German-Russian Collaboration                   
DFG-RFBR \\ \hspace*{3.5mm} (grant no. 436 RUS 113/248/3 and no. 436 RUS 113/248/2)\\              
% ...
                                                           %                                       
                                                           %                                       
% \par         % if index listing & table fit to 1 page, put gap here                              
\newpage   % alternatively: go to newpage, if page is too small                                    
                                                           %                                       
% \institute_references_start    % do not touch or move this line !                                
                                                           %                                       
\begin{tabular}[h]{rp{14cm}}                                                                       
$^{a}$ &  supported by the Natural Sciences and Engineering Research                               
          Council of Canada (NSERC)  \\                                                            
$^{b}$ &  supported by the FCAR of Qu\'ebec, Canada  \\                                            
$^{c}$ &  supported by the German Federal Ministry for Education and                               
          Science, Research and Technology (BMBF), under contract                                  
          numbers 057BN19P, 057FR19P, 057HH19P, 057HH29P, 057SI75I \\                              
$^{d}$ &  supported by the MINERVA Gesellschaft f\"ur Forschung GmbH, the                          
German Israeli Foundation, and by the Israel Ministry of Science \\                                
$^{e}$ &  supported by the German-Israeli Foundation, the Israel Science                           
          Foundation, the U.S.-Israel Binational Science Foundation, and by                        
          the Israel Ministry of Science \\                                                        
$^{f}$ &  supported by the Italian National Institute for Nuclear Physics                          
          (INFN) \\                                                                                
$^{g}$ &  supported by the Japanese Ministry of Education, Science and                             
          Culture (the Monbusho) and its grants for Scientific Research \\                         
$^{h}$ &  supported by the Korean Ministry of Education and Korea Science                          
          and Engineering Foundation  \\                                                           
$^{i}$ &  supported by the Netherlands Foundation for Research on                                  
          Matter (FOM) \\                                                                          
$^{j}$ &  supported by the Polish State Committee for Scientific Research,                         
          grant No. 115/E-343/SPUB/P03/154/98, 2P03B03216, 2P03B04616,                             
          2P03B10412, 2P03B05315, 2P03B03517, and by the German Federal                            
          Ministry of Education and Science, Research and Technology (BMBF) \\                     
$^{k}$ &  supported by the Polish State Committee for Scientific                                   
          Research (grant No. 2P03B08614 and 2P03B06116) \\                                        
$^{l}$ &  partially supported by the German Federal Ministry for                                   
          Education and Science, Research and Technology (BMBF)  \\                                
$^{m}$ &  supported by the Fund for Fundamental Research of Russian Ministry                       
          for Science and Edu\-cation and by the German Federal Ministry for                       
          Education and Science, Research and Technology (BMBF) \\                                 
$^{n}$ &  supported by the Spanish Ministry of Education                                           
          and Science through funds provided by CICYT \\                                           
$^{o}$ &  supported by the Particle Physics and                                                    
          Astronomy Research Council \\                                                            
$^{p}$ &  supported by the US Department of Energy \\                                              
$^{q}$ &  supported by the US National Science Foundation \\                                       
$^{r}$ &  partially supported by the British Council,                                              
          ARC Project 0867.00 \\                                                                   
$^{s}$ &  partially supported by the British Council,                                              
          Collaborative Research Project, TOK/880/11/15                                            
\end{tabular}                                                                                      
                                                           %                                       
% \institute_references_end     % do not touch or move this line !        

\newpage

% Set page numbering back to Arabic :
% ===================================
\pagenumbering{arabic}
\setcounter{page}{1}

\section{Introduction}

This paper reports the results of an investigation into the production of $W$ bosons in 
positron-proton collisions at HERA.  The collider operated from 1994 to 1997 with positron and 
proton beam energies of $27.5$ and $820$~${\rm GeV}$ respectively, resulting in a centre-of-mass 
energy of $300$~${\rm GeV}$. During this period the ZEUS detector collected data corresponding to an 
integrated luminosity of $47.7$~${\rm pb^{-1}}$.

The Standard Model calculation of the cross section for the production of $W$ bosons via the 
reaction $e^{+}p\to e^{+}W^{\pm}X$ yields a value of roughly 
$1$~${\rm pb}$~\cite{baur89, baur92, dubinin}. The $W$-production cross section is sensitive 
to the couplings at the $WW\gamma$ vertex, particularly at large hadronic transverse 
momentum (``hadronic $P_{T}$''). A measurement of the cross section can therefore provide limits 
on anomalous $WW\gamma$ couplings. In addition, the measurement gives a useful constraint on 
the $W$-production background to a variety of searches for non-Standard-Model physics at HERA.

The search for signals for $W$ production in both the electron\footnote{The term electron 
is used to refer to both electrons and positrons.} and muon decay channels was performed by
selecting events with large missing transverse momentum (``missing $P_{T}$'') which also 
contain isolated leptons with high transverse momentum (``high $P_{T}$''). The result of the 
search in each decay channel is consistent with Standard Model expectations. The integrated
luminosity and the signal-to-background ratio in the electron channel search are sufficient 
to allow a first estimate of the cross section for $W$ production in electron-proton 
interactions. The searches in the two decay channels are combined at large hadronic $P_{T}$ in order to obtain cross-section limits in this region, and 
limits on the couplings at the $WW\gamma$ triple-gauge-boson vertex are calculated.

The H1 collaboration has recently reported the observation of six events containing isolated high-energy
leptons and missing transverse momentum~\cite{h1_paper}. The number of muon events is significantly 
larger than the Standard Model expectation, to which $W$ production forms the major contribution. A search 
for such events with the ZEUS detector, using similar cuts and with a similar sensitivity to that of the H1 
analysis, yields results in good agreement with the Standard Model.

In Section~2 of this article, the $W$-production signal and its simulation are discussed in more detail. 
Background processes are treated in Section~3. Section~4 describes the ZEUS detector. Details of the
event reconstruction and pre-selection are given in Section~5. The results of the analysis in the 
electron and muon channels are presented in Sections~6 and~7, respectively. Section~8 presents upper limits on the cross section for $W$ production at 
large hadronic $P_{T}$, while Section~9 presents the ZEUS analysis of events containing an isolated 
high-energy charged particle in addition to large missing transverse momentum. The results are 
summarised in Section~10.

\section{$W$ Production at HERA}
\label{sec:w_production}

The dominant $W$-production process in $e^{+}p$ collisions at HERA is the reaction
\begin{equation}
e^{+}p\to e^{+}W^{\pm}X\:\:,  \label{reaction:epeWX}
\end{equation}
in which the scattered beam electron emerges at small angles with respect to the lepton beam direction 
and is generally not found in the central detector. The observed event topology therefore consists of
the hadronic final state $X$, which typically carries small transverse momentum, and the $W$ decay 
products at comparatively large laboratory angles.

\subsection{Cross-Section Calculation}

\begin{figure}[!t]
 \begin{center}
  \mbox{\epsfig{figure=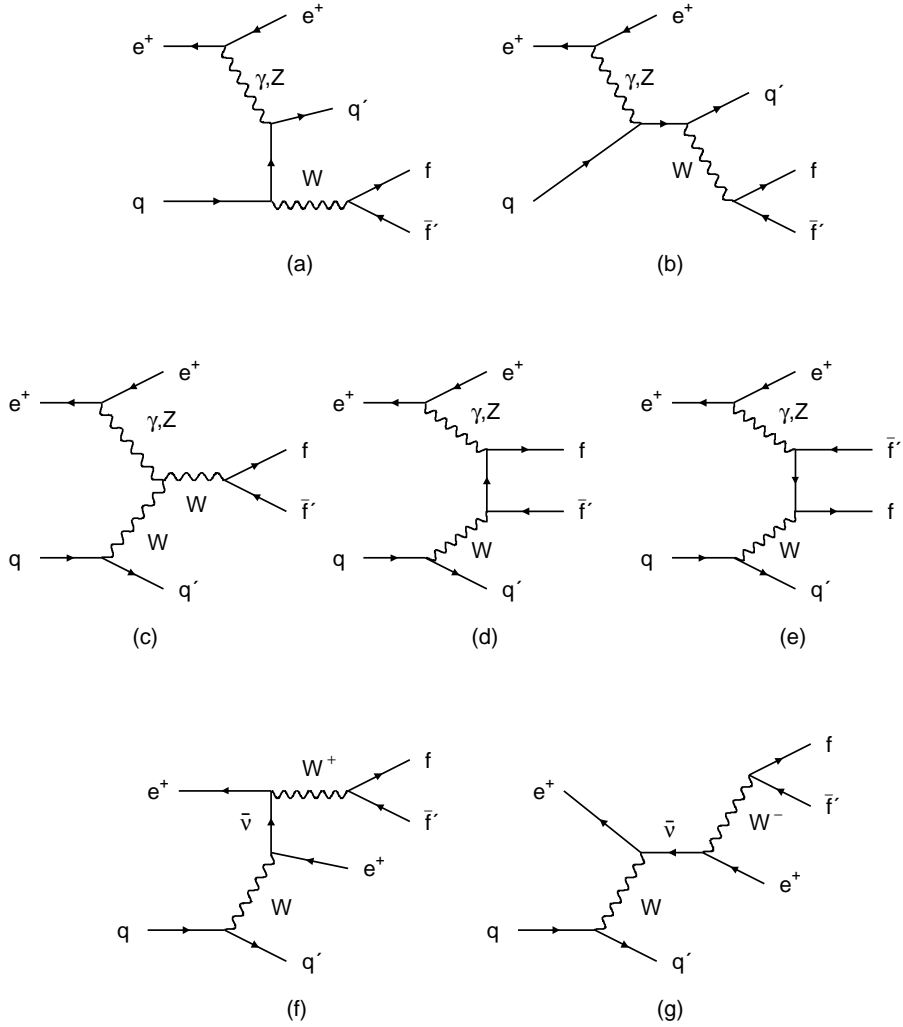,width=0.8\textwidth}}
 \end{center}
\caption{{\em Leading-order Feynman diagrams for the process $e^{+}p\to e^{+}W^{\pm}X, W\to{f{\bar f}'}$. 
See text for more details.}}
\label{fig:w_feyn}
\end{figure}

The leading-order diagrams for reaction~(\ref{reaction:epeWX}) are shown in Fig.~\ref{fig:w_feyn}. 
Diagrams (a) and (b) correspond to $W$ radiation from the incoming and scattered quark, respectively.
Diagrams (f) and (g), in which the $W$ couples to the incoming or scattered lepton line, are suppressed 
by a second heavy propagator. Diagram (c) contains the $WW\gamma$ triple-gauge-boson coupling. 
Diagrams (d) and (e), required to preserve gauge invariance, contain off-shell $W$'s which give rise 
primarily to low-$P_{T}$ charged leptons and lepton-neutrino invariant masses far from the $W$ mass.

The contributions of the different diagrams are calculated with the Monte Carlo based program 
{\sc EPVEC}~\cite{baur92}. The fermion $u$-channel pole of diagram (a) is regularised by splitting 
the phase space into two regions :
\begin{displaymath}
\sigma = \sigma (|u|> u_{cut}) + \int^{u_{cut}} \frac{d\sigma}{d|u|} \:\: d|u|
\end{displaymath}
where $u=(p_{q}-p_{W})^{2}$ and $p_{q}$, $p_{W}$ are the four momenta of the incoming quark 
and final state $W$ boson, respectively. The first term is calculated using helicity amplitudes 
for the process $e^{+}q\to e^{+}Wq',W\to{f{\bar f}'}$. The cross section for small values of $|u|$ is 
calculated by folding the cross section for $q\bar q'\to W\to {f{\bar f}'}$ with the parton densities
in the proton and the effective parton densities for the resolved photon emitted by the incoming electron.
The resulting total cross section for reaction~(\ref{reaction:epeWX}) varies little with $u_{cut}$, 
chosen here to be $25$~${\rm GeV^2}$. 

Using the MRS(G)~\cite{mrsg} set of parton densities in the proton evaluated at a scale $M_{W}^{2}$ and the
GRVG-LO~\cite{GRVG-LO} set of parton densities in the photon evaluated at a scale $p_{W}^{2}/10$, the 
cross sections are $0.52$~${\rm pb}$ for $W^{+}$ and $0.42$~${\rm pb}$ for $W^{-}$ production
via reaction~(\ref{reaction:epeWX}), giving a total of $0.95$~${\rm pb}$.

The cross section for the process $e^{+}p\to\bar\nu W^{+}X$, also calculated using {\sc EPVEC}, 
is only about $5\%$ of that for reaction~(\ref{reaction:epeWX}). The $Z^{0}$ production process 
$e^{+}p\to e^{+}Z^{0}X$, with a cross section of around $0.3$~${\rm pb}$, has been simulated using 
{\sc EPVEC} in order to estimate the contribution of $Z^{0}\to l^{+}l^{-}$ and $Z^{0}\to\nu\bar\nu$ 
decays to the high-$P_{T}$ lepton samples considered here.

\subsection{Cross-Section Uncertainties}

The use of different proton and photon parton densities changes the calculated $W$-production
cross section by up to $5\%$ and $10\%$, respectively. Large uncertainties also result from the choice
of hard scale used to evaluate the structure functions. Added in quadrature, the combined effect of 
these uncertainties leads to an estimated overall uncertainty in the $W$-production cross section of 
about $20\%$. 

{\sc EPVEC} is a leading-order program and includes no QCD radiation. A recent calculation of the cross
section for reaction~(\ref{reaction:epeWX}) includes a next-to-leading-order (NLO) calculation of the resolved-photon 
contribution~\cite{spira}. The result is $0.97$~${\rm pb}$, close to the {\sc EPVEC} estimate 
of $0.95$~${\rm pb}$. The scale dependence of the NLO cross section is reduced to the $5$-$10\%$ level, 
although the structure-function-related uncertainties remain. Changes to the hadronic-$P_{T}$ spectrum 
due to higher-order effects, which could be important for some acceptance calculations and for setting 
coupling limits, are currently unknown.

\section{Background Processes}
\label{sec:mc}

The most important background to $W$ production in the electron decay channel arises from 
high-$Q^{2}$ charged- and neutral-current deep inelastic scattering.  These have 
both been simulated using the event generator {\sc DJANGO6}~\cite{django}, an interface 
to the Monte Carlo programs {\sc HERACLES~4.5}~\cite{heracles} and {\sc LEPTO~6.5}~\cite{lepto}. 
Leading-order QCD and electroweak radiative corrections are included and higher-order QCD effects 
are simulated via parton cascades using both the parton shower and matrix elements approach of {\sc LEPTO} 
and the colour-dipole model {\sc ARIADNE}~\cite{ariadne}. The final hadronisation of the partonic final 
state is performed with {\sc JETSET}~\cite{jetset}.

Two-photon processes provide an additional source of high-$P_{T}$ leptons which are a significant background 
in the muon decay channel.  The dominant, Bethe-Heitler, process has been simulated using the event 
generator {\sc LPAIR}~\cite{lpair} including both elastic and inelastic production at the proton vertex.

Finally, photoproduction has been simulated with the {\sc HERWIG}~\cite{herwig} Monte Carlo program, 
including both resolved and direct photon contributions.

\section{The ZEUS Detector}

A detailed description of the ZEUS detector can be found elsewhere~\cite{zeus_1,zeus_2}. The main 
components used in this analysis were the compensating uranium-scintillator calorimeter (CAL) and the 
central tracking detector (CTD).

The CAL is divided into three parts, forward (FCAL) covering the polar angle\footnote{The ZEUS 
coordinate system is right-handed with the $Z$-axis pointing in the proton beam direction and the 
horizontal $X$-axis pointing towards the centre of HERA. The polar angle, $\theta$, is measured with 
respect to the $+Z$-axis and the pseudorapidity, $\eta$, is related to the polar angle 
by $\eta=-\ln(\tan(\theta/2))$.} interval $3^{\circ}<\theta<37^{\circ}$, barrel (BCAL) covering the 
range $37^{\circ}<\theta<129^{\circ}$ and rear (RCAL) covering the range 
$129^{\circ}<\theta<176^{\circ}$, as viewed from the nominal interaction point~\cite{ucal}. 
Each part is divided into towers approximately $20\times 20$~${\rm cm}$ in transverse size and 
segmented longitudinally into an electromagnetic (EMC) section and two hadronic (HAC) sections 
(one in RCAL).  Within the EMC section each tower is further subdivided transversely into four cells 
(two in RCAL).  Each cell is read out by a pair of wavelength shifters and photomultiplier tubes.  
Calorimeter energy resolutions of $\sigma_{E}/E=18\%/\sqrt{E({\rm GeV})}$ for electrons and 
$\sigma_{E}/E=35\%/\sqrt{E({\rm GeV})}$ for hadrons have been measured under test-beam conditions.
An instrumented-iron backing calorimeter (BAC) measures energy leakages from the central
uranium calorimeter~\cite{bac}.

The CTD is a cylindrical multi-wire drift chamber operating in a $1.43$~${\rm T}$ solenoidal magnetic 
field~\cite{ctd}. A momentum measurement, for tracks passing through at least 2 of the 9 radial 
superlayers, can be made in the polar angle range $15^{\circ}<\theta<164^{\circ}$.  \linebreak The 
transverse-momentum resolution for full length tracks can be parameterised 
as \linebreak $\sigma(P_{T})/P_{T}= 0.0058\;P_{T}\oplus 0.0065 \oplus 0.0014/P_{T}\;$, with $P_{T}$ in ${\rm GeV}$.

The luminosity is determined from the rate of high-energy photons produced in the process $ep\to ep\gamma$ 
which are measured in a lead-scintillator calorimeter located at \linebreak $Z=-107$~${\rm m}$~\cite{lumi}. 

The ZEUS three-level trigger system efficiently selects events with large missing and total transverse 
energies~\cite{zeus_2}. Several triggers at each level are used to tag events used for this analysis, 
with the relevant energy thresholds generally reduced if a good CTD track is present in addition to large 
calorimeter energies. Algorithms based on tracking and calorimeter timing information reject non-$ep$ 
backgrounds, consisting mainly of proton beam-gas interactions and cosmic rays.

\section{Event Reconstruction and Pre-selection}
\label{sec:presel}

The calorimeter transverse momentum is defined as :
\begin{equation}
\mbox{calorimeter } P_{T}\;=\;\sqrt{(\sum_{i}p_{X,i})^{2}+(\sum_{i}p_{Y,i})^{2}}\:\:, 
\end{equation}
where $p_{X,i}=E_{i}\sin{\theta_{i}}\cos{\phi_{i}}$ and $p_{Y,i}=E_{i}\sin{\theta_{i}}\sin{\phi_{i}}$ 
are calculated using the energies ($E_{i}$) of individual calorimeter cells that are above noise thresholds of 
$80$~${\rm MeV}$ (EMC) and $140$~${\rm MeV}$ (HAC). The angles $\theta_{i}$ and $\phi_{i}$ are estimated from 
the geometric cell centres and the event vertex. Note that in $W\to e\nu$ events, calorimeter $P_{T}$ as defined above 
is an estimate of the missing $P_{T}$ or transverse momentum carried by the neutrino. In muon events, a combination 
of the calorimeter $P_{T}$ and the transverse momentum of the muon track measured in the CTD
is used to calculate the missing $P_{T}$.

Electron (hadron) transverse momenta are defined as sums over those calorimeter cells that are (are not) assigned 
to the electron candidate cluster. Longitudinal momentum conservation ensures that $E-p_{Z}$, 
defined as : 
\begin{displaymath}
E-p_{Z}=\sum_{i}{E_{i}(1-\cos{\theta_{i}})}\:\:,
\end{displaymath}
peaks at $2E_{e}$ for fully contained events, where $E_{e}$ is the electron beam energy. 
Smaller values of $E-p_{Z}$ indicate energy escaping detection, either in the rear beam pipe 
or in the form of muons or neutrinos.

\begin{figure}[!tb]
 \begin{center}
  \mbox{\epsfig{figure=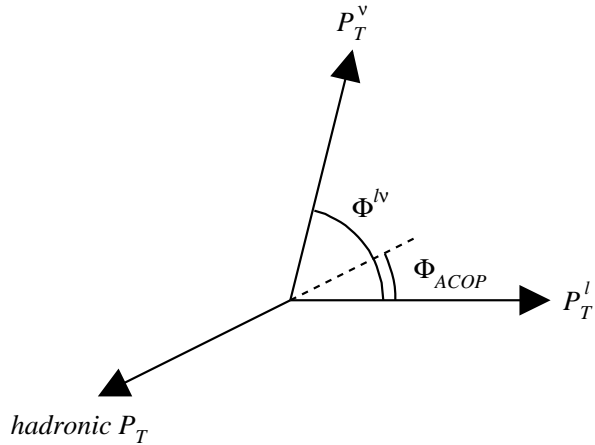,width=0.5\textwidth}}
 \end{center}
\caption{{\em The definition of the acoplanarity angle, $\Phi_{\rm ACOP}$, and the azimuthal separation 
of the neutrino and outgoing lepton, $\Phi^{l\nu}$, in the transverse plane. The dashed line balances the 
hadronic $P_{T}$. See text for more details.}}
\label{fig:acop_paper}
\end{figure}

The acoplanarity angle $\Phi_{\rm ACOP}$, illustrated in Fig.~\ref{fig:acop_paper}, is the
azimuthal separation of the outgoing lepton and the vector in the $\{X,Y\}$-plane that balances
the hadronic-$P_{T}$ vector. For well measured neutral-current events the acoplanarity 
angle is close to zero, while large acoplanarity angles indicate large missing energies.  
The transverse mass~\cite{ua1_mt} is defined as 
\begin{displaymath}
M_{T}=\sqrt{2 \: P_{T}^{l} \: P_{T}^{\nu} \: (1-\cos\Phi^{l\nu})}\:\:,
\end{displaymath}
where $P_{T}^{l}$ is the lepton transverse momentum, $P_{T}^{\nu}$ is the magnitude of the missing~$P_{T}$ 
and $\Phi^{l\nu}$ is the azimuthal separation of the lepton and missing-$P_{T}$ vectors, as
shown in Fig.~\ref{fig:acop_paper}.

Events that pass the trigger requirements are further required to have a reconstructed calorimeter $P_{T}$ 
in excess of $12$~${\rm GeV}$. The transverse momentum, calculated excluding the inner ring of
calorimeter cells around the forward beam-pipe hole, must be greater than $9$~${\rm GeV}$. 
These offline cuts are more stringent than the corresponding online trigger thresholds in any given year 
of data taking. Other pre-selection cuts common to both the electron and muon event selections are the 
requirements that the $Z$-coordinate of the tracking vertex be reconstructed within $50$~${\rm cm}$ 
of the nominal interaction point and have at least one associated track with transverse momentum greater 
than $0.2$~${\rm GeV}$ and a polar angle in the range $15^{\circ}<\theta<164^{\circ}$.  Cuts on the calorimeter 
timing and algorithms based on the pattern of tracks in the CTD reject beam-gas, cosmic-ray and halo-muon events.

\section{Search for $W$ Production and Decay $W\to e\nu$}

Electron-identification criteria are applied to the pre-selected events and the data are subsequently 
compared to the Monte Carlo simulation. Final results are presented after further cuts designed to 
optimise the sensitivity to the $W$-production signal.

\subsection{Electron Identification}

A neural-network-based algorithm to identify electrons, trained on Monte Carlo events and optimised for maximum 
electron-finding efficiency and electron-hadron separation, selects candidate electromagnetic clusters in the 
calorimeter~\cite{sinistra}. A cut on the electromagnetic-cluster energy of $8$~${\rm GeV}$ is made, above
which the neural network is fully efficient except at the boundaries between the different calorimeter parts.
The impact point of the electron at the face of the calorimeter is determined with a resolution of $1$~${\rm cm}$ using the pulse 
height information from the pairs of photomultipliers reading out each cell. The distance of closest approach 
of a matching extrapolated CTD track to the electromagnetic cluster is required to be less than $10$~${\rm cm}$, 
where only tracks with $15^{\circ}<\theta<164^{\circ}$ are considered. The background from fake electrons is reduced by 
requiring that the energy not associated with the electron in an $\{\eta,\phi\}$ cone of radius $0.8$ around the 
electron direction be less than $4$~${\rm GeV}$. Moreover, since most fake electron candidates are 
misidentified hadrons close to jets, this background is further reduced by requiring that the electron track 
be separated by at least $0.5$ units in $\{\eta,\phi\}$ space from other tracks associated with the event vertex.

\subsection{Comparison of Data and Monte Carlo}

\begin{figure}[p]
 \begin{center}
  \mbox{\epsfig{figure=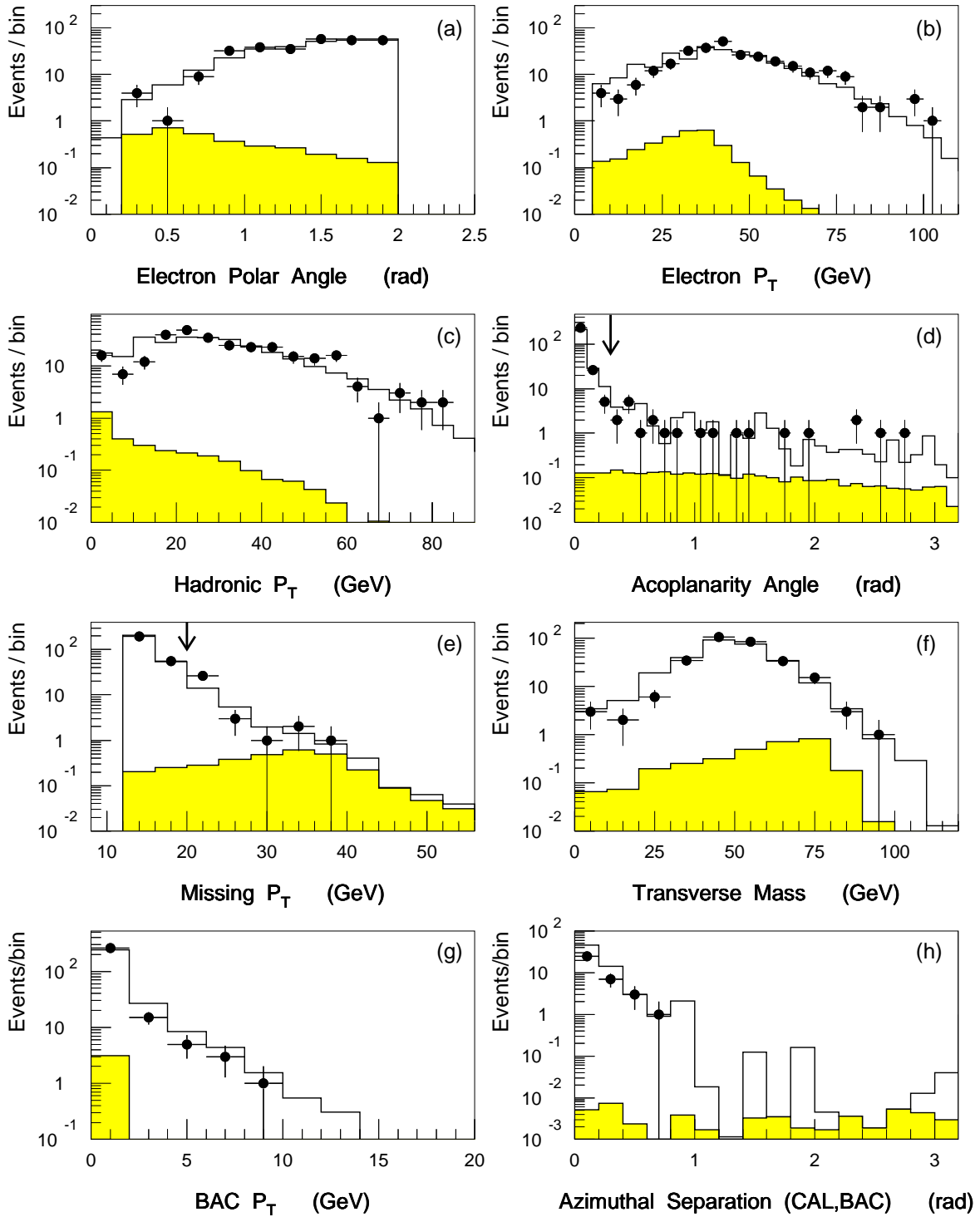,width=1.0\textwidth}}
 \end{center}
%crucial backspace to avoid text/pagenumber conflict!
\vskip -0.5cm
\caption{{\em The data (points) compared to the combined luminosity-weighted Monte Carlo expectation 
(open histograms) for events containing a high-$P_{T}$ electron in addition to large missing~$P_{T}$. 
The $W$-production component of the Monte Carlo is indicated by the shaded histogram for each distribution.
The cuts on the acoplanarity angle and missing $P_{T}$ used to suppress the neutral-current background
are indicated by the arrows in (d) and (e). Only events with BAC energy deposits are plotted in (h).}}
\label{fig:pt_e_presel}
\end{figure}

The data are compared to the expectation from the Monte Carlo simulation in Fig.~\ref{fig:pt_e_presel}, 
after requiring that the transverse momentum of the electron, $P_{T}^{e}$, be greater than $5$~${\rm GeV}$ 
and the polar angle of the electron measured in the calorimeter, $\theta_{e}$, be less than 
$2.0$~${\rm rad}$. Neutral-current background events dominate the sample at 
this stage of the selection, as is evident from the steeply falling missing-$P_{T}$ spectrum and the concentration of 
events at small acoplanarity angles. A Jacobian-peak structure is visible in the transverse-mass distribution 
for the Monte Carlo simulation of the signal events, shown in Fig.~\ref{fig:pt_e_presel}(f). 
Figure~\ref{fig:pt_e_presel}(g) shows the $P_{T}$ reconstructed in the backing calorimeter and 
Fig.~\ref{fig:pt_e_presel}(h) shows the azimuthal separation between the BAC $P_{T}$ and uranium-calorimeter 
missing-$P_{T}$ directions, for events with BAC energy deposits. All distributions show reasonable agreement 
between the data and Monte Carlo. 

\subsection{Final Cuts and Results} \label{sec:w_e_final}

The neutral-current background is heavily suppressed by requiring the missing~$P_{T}$ to be greater than 
$20$~${\rm GeV}$ and the acoplanarity angle to be greater than $0.3$~${\rm rad}$, indicated by the arrows in 
Figs.~\ref{fig:pt_e_presel}(d) and (e). The latter cut is only applied to events with a hadronic~$P_{T}$ 
in excess of $4$~${\rm GeV}$, for which the acoplanarity angle is well defined. Electrons in the final event sample 
must, in addition, have $P_{T}^{e}>10$~${\rm GeV}$ and $\theta_{e}<1.5$~${\rm rad}$. Neutral-current background is 
further reduced by removing events with energy deposits in the backing calorimeter that are closely aligned with 
the direction of the missing~$P_{T}$.  Finally, requiring that the matching electron track have a transverse 
momentum greater than $5$~${\rm GeV}$, as measured in the CTD, removes most of the remaining fake electrons.

\begin{figure}[!tb]
 \begin{center}
  \mbox{\epsfig{figure=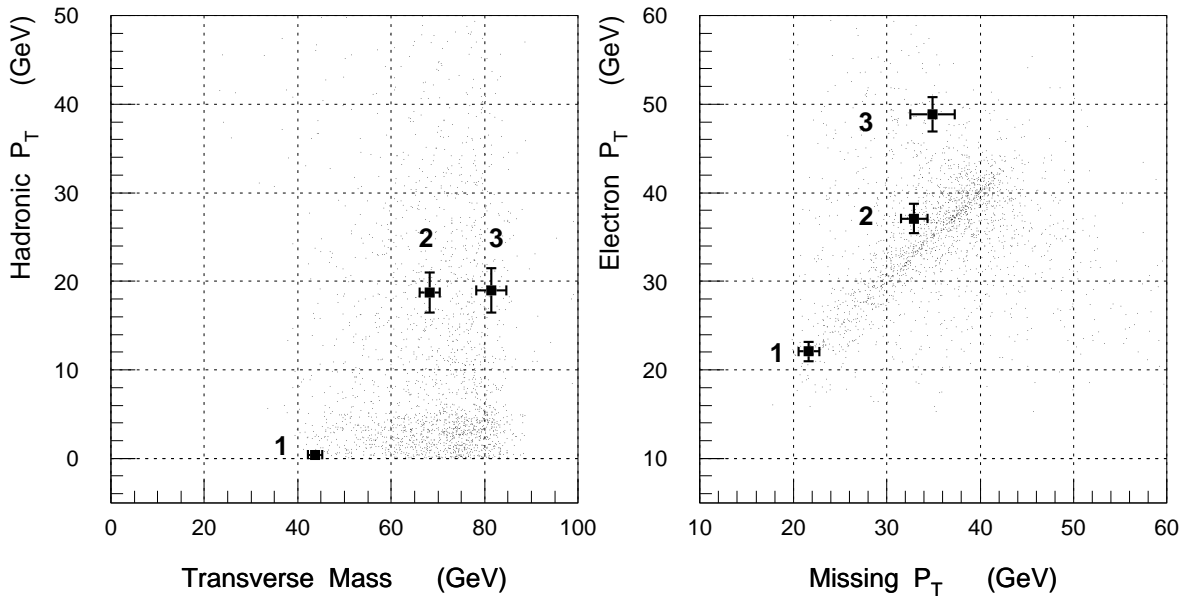,width=1.0\textwidth}}
 \end{center}
\caption{{\em Characteristic kinematic variables of the three surviving events in the electron sample (square points) and 
$W$-production Monte Carlo corresponding to $50$~${\rm fb^{-1}}$ (dots) after all cuts.  All quantities are 
corrected as described in the text and the errors on the data points are indicated by the error bars.
Note that the Monte Carlo simulation is not expected to reproduce accurately the true hadronic-$P_{T}$ distribution 
in the region of low hadronic $P_{T}$, where $W$ production via resolved photons dominates.}}
\label{fig:w_e_scatter}
\end{figure}

\begin{figure}[!tb]
 \begin{center}
  \mbox{\epsfig{figure=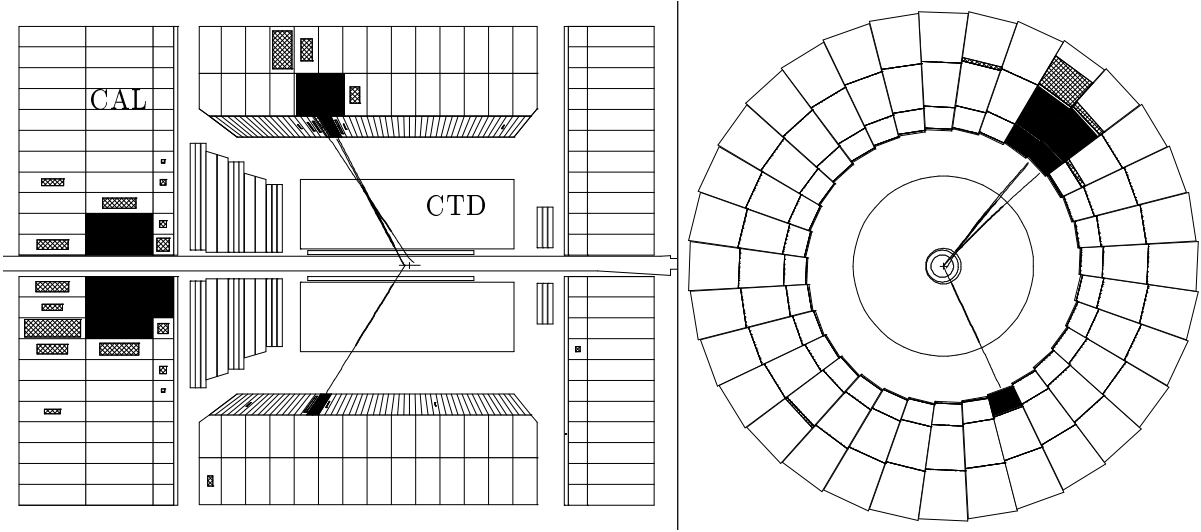,width=1.0\textwidth,clip=}}
 \end{center}
\caption{{\em Event~3 in the final $W\to e\nu$ data sample, shown in a plane parallel (left) and 
perpendicular (right) to the beam line. The shaded areas in the calorimeter indicate energy deposits.
An electron is visible in the lower half of the segmented uranium calorimeter, acoplanar with a 
jet of hadrons visible in the upper half. The transverse mass for the event is $81.4$~$\pm$~$3.2$~${\rm GeV}$.}}
\label{fig:w_e_evt_pic}
\end{figure}

Three data events, all of which have a final-state $e^{+}$, survive these cuts. The properties of these events 
are given in Table~\ref{tab:w_e_properties} and compared in Fig.~\ref{fig:w_e_scatter} 
to the $W$-production Monte Carlo with all cuts applied. 
For these figures and the table, the electron and hadron energies used in calculating the missing~$P_{T}$ 
and transverse mass have been corrected for the effect of the inactive material between the $ep$ interaction 
point and the calorimeter. The corrections are typically a few percent for the electron and $10\%$ for the 
hadron transverse momenta. The momenta of the electrons measured in the CTD agree within errors with the 
corrected associated calorimeter energies. Given the different cross sections and selection efficiencies for $W^{+}$ and 
$W^{-}$ production and decay, roughly $60\%$ of signal events are expected to have an $e^{+}$ rather than an $e^{-}$ 
in the final state, while background events are expected to contain predominantly $e^{+}$ candidates.
The charge composition of the sample is therefore consistent with expectations. 
The event in Table~\ref{tab:w_e_properties} with the highest transverse mass is illustrated in Fig.~\ref{fig:w_e_evt_pic}. 
Event~2 has a similar topology while event~1 has no visible hadronic jet and consequently a small value for the reconstructed 
hadronic~$P_{T}$.
\begin{table}[t]
\renewcommand{\arraystretch}{1.5}
\begin{center}
\begin{tabular}{|c||c|c|c|}
\hline
\multicolumn{1}{|c||}{candidate} & \multicolumn{1}{|c|}{1} & \multicolumn{1}{|c|}{2} & 
\multicolumn{1}{|c|}{3} \\ \hline \hline
electron polar angle & $37^{\circ}$ & $54^{\circ}$ & $58^{\circ}$ \\ \hline
$P_{T}^{e}$ (${\rm GeV}$) & $\:\:\:22.1$ $\pm$ $1.1\:\:\:$ & 
$\:\:\:37.1$ $\pm$ $1.6\:\:\:$ & $\:\:\:48.8$ $\pm$ $2.0\:\:\:$ \\ \hline
matching track $P_{T}$ (${\rm GeV}$) & $22.3^{+6.8}_{-4.2}$ & 
$35.9^{+13.8}_{-7.8}$ & $44.1^{+33.0}_{-13.2}$ \\ \hline
track charge & $+1\;(>4\,\sigma)$ & $+1\;(>3\,\sigma)$ & $+1\;(>2\,\sigma)$ \\ \hline
hadronic $P_{T}$ (${\rm GeV}$) & $0.4$ $\pm$ $0.1$ & 
$18.7$ $\pm$ $2.2$ & $19.0$  $\pm$ $2.5$ \\ \hline
hadronic $E-p_{Z}$ (${\rm GeV}$) & $0.50$ & $19.4$ & $22.0$ \\ \hline
missing $P_{T}$ (${\rm GeV}$) & $21.7$ $\pm$ $1.1$ & $32.9$ $\pm$ $1.4$ & $34.9$ $\pm$ $2.4$ \\ \hline
transverse mass (${\rm GeV}$) & $43.7$ $\pm$ $1.6$ & $68.2$ $\pm$ $2.2$ & $81.4$ $\pm$ $3.2$ \\ \hline
\end{tabular}
\end{center}
\caption{{\em The properties of the three surviving events in the search for $W$ production and 
decay $W\to e\nu$. The numbers in parentheses after the track charge indicate the significance of the 
sign determination.}}
\label{tab:w_e_properties}
\vspace{1.0cm}
\end{table}

The Monte Carlo expectation after all cuts is $2.1$ signal events and $1.1$~$\pm$~$0.3$ background 
events. The background consists mainly of charged-current DIS, with smaller 
contributions from Bethe-Heitler di-lepton production and $Z^{0}\to\nu\bar\nu$ events in which the 
beam electron is scattered at small polar angles. Taking into account the efficiency for selecting 
$W\to e\nu$ events of $38\%$ and the small extra sensitivity resulting from a $\sim 2\%$ efficiency 
for selecting $W\to\tau\nu$ decays, the three events, after subtracting the expected background, 
correspond to a measured cross section for reaction~(\ref{reaction:epeWX}) of :
\begin{displaymath}
\sigma (e^{+}p\to e^{+}W^{\pm}X) \:\:=\:\:\: 0.9 \:^{\:+\:1.0}_{\:-\:0.7} \:\:\:\mbox{(stat.)}\:\: \pm \:\: 
0.2 \:\:\:\mbox{(syst.)}\:\:\: {\rm pb}\:\:\:. 
\end{displaymath}
The systematic error is a combination of the uncertainty in the selection efficiency for $W\to e\nu$ events, 
the uncertainty in the estimate of the remaining background to $W$ production and decay, and a small 
contribution from the uncertainty in the total integrated luminosity. The systematic uncertainty in the 
selection efficiency arises from uncertainties in the electron-finding procedure and the $3\%$ uncertainty 
in the absolute calorimeter energy scale. The error on the background estimate is a combination of 
statistical errors and uncertainties due to the choice of model for simulating parton cascades in the DIS Monte Carlo.

 From the three observed events, an explicit upper limit on the $W$-production cross section has been 
derived, using the background expected from Monte Carlo and applying the method described in~\cite{pdg} :
\begin{displaymath}
\sigma (e^{+}p\to e^{+}W^{\pm}X) \:\:<\:\: 3.3 \:\:\: {\rm pb}\:\: 
\rm{at~95\%~C.L.}
\end{displaymath}
The selection efficiency in the electron channel depends little on the recoiling hadronic~$P_{T}$. This implies
that the upper limit given above is insensitive to uncertainties in the underlying hadronic-$P_{T}$ distribution 
arising from higher-order effects or anomalous $WW\gamma$ couplings.

\section{Search for $W$ Production and Decay $W\to\mu\nu$}

The search for reaction~(\ref{reaction:epeWX}) with the subsequent decay $W\to\mu\nu$ 
begins with the same sample of events with a large calorimeter missing~$P_{T}$ used in the 
electron analysis (see Section~\ref{sec:presel}). However, since a high-energy muon leaves 
only a small energy deposit in the calorimeter, this selection necessarily restricts the 
acceptance to $W$-production events with large hadronic transverse momenta.

\subsection{Muon Identification}

The energy deposited by minimum-ionising particles (MIP's) can be distributed across several 
calorimeter clusters. Therefore neighbouring clusters are grouped together into larger scale objects which, 
providing they pass topological and energy cuts, are called calorimeter MIP's. In this analysis, a muon 
candidate is simply a calorimeter MIP that matches an extrapolated CTD track within $20$~${\rm cm}$, where
only tracks in the polar angle range $15^{\circ}<\theta<164^{\circ}$ are considered.  
The muon transverse momentum, $P_{T}^{\mu}$, and direction, including the polar angle, $\theta_{\mu}$, are 
obtained from the matching CTD track. The same energy- and track-isolation requirements made in the electron 
analysis are also applied here to the muon candidate.

\subsection{Comparison of Data and Monte Carlo}

\begin{figure}[!tb]
 \begin{center}
  \mbox{\epsfig{figure=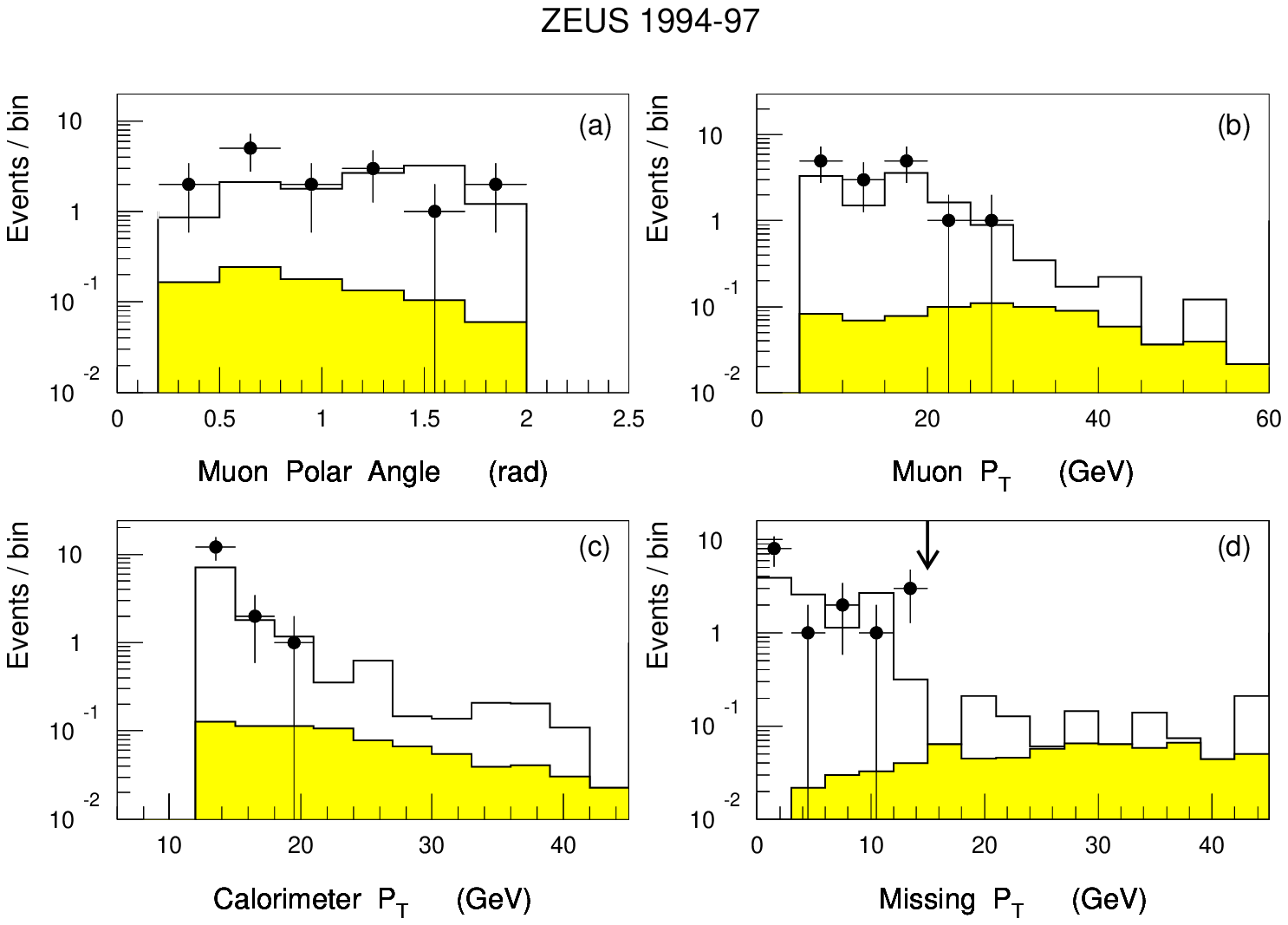,width=1.0\textwidth}}
 \end{center}
\caption{{\em The data (points) compared to the combined luminosity-weighted Monte Carlo expectation 
(open histogram) for events containing a high-$P_{T}$ muon in addition to large calorimeter~$P_{T}$. 
The $W$-production component of the Monte Carlo is given in each distribution by the shaded histogram.
The final cut on the missing $P_{T}$ is indicated by the arrow in (d).}}
\label{fig:pt_mu_presel}
\end{figure}

The data and Monte Carlo predictions are compared in Fig.~\ref{fig:pt_mu_presel}, after requiring 
$P_{T}^{\mu}>5$~${\rm GeV}$ and $\theta_{\mu}<2.0$~${\rm rad}$. Events with more than one muon 
candidate having $P_{T}^{\mu}>2$~${\rm GeV}$ have been removed.  Fig.~\ref{fig:pt_mu_presel}(d) shows the 
missing transverse momentum, calculated by combining the muon and calorimeter $P_{T}$'s in the transverse
plane after subtracting the muon's contribution to the latter. In each case the 
distribution of events is similar to that expected from the Monte Carlo simulation of the background, 
which is dominated by Bethe-Heitler di-muon production.

\subsection{Final Cuts and Results} \label{sec:w_mu_final}

The final stage in the selection of $W\to\mu\nu$ events requires the missing transverse 
momentum to be greater than $15$~${\rm GeV}$. No data event survives this final cut, shown in
Fig.~\ref{fig:pt_mu_presel}(d), to be compared with an expected $0.76$ events from $W$ production and 
$0.65\pm 0.22$ from background. The latter is dominated by charged-current DIS and 
Bethe-Heitler~$\mu^{+}\mu^{-}$ production. The efficiency for selecting $W\to\mu\nu$ 
events is $13\%$, lower than the corresponding efficiency in the electron channel due to the soft 
hadronic-$P_{T}$ spectrum expected for reaction~(\ref{reaction:epeWX}). The resulting $95\%$~C.L. 
upper limit on the cross section for this reaction is therefore weaker, at $3.7$~${\rm pb}$. Note that 
the value for the efficiency in the muon channel, calculated here using {\sc EPVEC}, is much more
sensitive to assumptions about the hadronic-$P_{T}$ distribution than the corresponding efficiency
in the electron channel.

\section{$W$ Production at Large Hadronic~$P_{T}$}

The final electron and muon samples described in Sections~\ref{sec:w_e_final} and~\ref{sec:w_mu_final}, 
respectively, are combined and events which have large hadronic $P_{T}$ are selected. The number of 
events with a corrected hadronic~$P_{T}$ above specified values is shown in Table~\ref{tab:had_pt_cuts}, along 
with the number of $W$-production and background events expected from Monte Carlo. The efficiencies listed in 
the table have been calculated for the subset of $W$-production events that have a true hadronic~$P_{T}$ above 
the given value. They are averaged over all decay channels, thereby including a small contribution from $W\to\tau\nu$
decays. The resulting $95\%$~C.L. upper limits on the cross section for the reaction $e^{+}p\to e^{+}W^{\pm}X$ 
at large hadronic $P_{T}$ are also given in the table.
Note that the selection efficiency in 
the muon channel is small for low hadronic~$P_{T}$, reaching a plateau comparable to the efficiency in the electron 
channel at around $20$~${\rm GeV}$. Cross-section limits for cuts on the hadronic $P_{T}$ at or above this
value are insensitive to the underlying hadronic-$P_{T}$ distribution.
\renewcommand{\arraystretch}{1.5}
\begin{table}[t]
\begin{center}
\begin{tabular}{|c||c||c|c|c||c|}
\hline
\multicolumn{1}{|c||}{} &
\multicolumn{1}{|c||}{} &
\multicolumn{3}{|c||}{Standard Model Monte Carlo} &
% \multicolumn{1}{|c||}{W} &
\multicolumn{1}{|c|}{$95\%$ C.L. cross section} \\ \cline{3-5}
\multicolumn{1}{|c||}{\raisebox{0.3cm}{hadronic $P_{T}$}} &
\multicolumn{1}{|c||}{\raisebox{0.3cm}{data}} &
\multicolumn{1}{|c|}{W } &
\multicolumn{1}{|c|}{background } &
\multicolumn{1}{|c||}{W efficiency} &
\multicolumn{1}{|c|}{upper limit} \\ \hline \hline
$>$ $15$ ${\rm GeV}$ & $2$ & $1.4$ & $0.62$~$\pm$~$0.20$ & $10\%$  & $1.2$~${\rm pb}$ \\ \hline
$>$ $20$ ${\rm GeV}$ & $0$ & $1.2$ & $0.52$~$\pm$~$0.18$ & $11\%$  & $0.58$~${\rm pb}$ \\ \hline
$>$ $25$ ${\rm GeV}$ & $0$ & $0.89$ & $0.49$~$\pm$~$0.18$ & $11\%$  & $0.56$~${\rm pb}$ \\ \hline
\end{tabular}
\end{center}
\caption{{\em The results of a search for events with large hadronic~$P_{T}$ in the combined
electron and muon samples. The number of observed events is compared to the Monte Carlo
expectation for Standard Model $W$ production and background,
after a cut on the corrected hadronic~$P_{T}$.  Selection efficiencies are averaged over all $W$
decay channels and the upper limits are on the cross section for reaction~(\ref{reaction:epeWX})
above the indicated hadronic~$P_{T}$.}}
\label{tab:had_pt_cuts}
\end{table}

The limits given in Table~\ref{tab:had_pt_cuts} can be used to constrain various any new physics processes 
which produce events with a $W$ boson and large hadronic~$P_{T}$. Such model-dependent 
analyses are outside of the scope of this paper. It is nevertheless useful to parameterise 
such effects in terms of anomalous $WW\gamma$ couplings,
which give rise to a harder distribution of the transverse momentum of the $W$ 
than expected in the Standard Model.
The most general effective
Lagrangian that is consistent with Lorentz invariance, CP conservation and electromagnetic gauge invariance, 
has two free couplings at the $WW\gamma$ vertex that are conventionally labelled $\kappa$ and 
$\lambda$~\cite{baur89,hagiwara}. In the Standard Model they take the values $\kappa=1$ and $\lambda=0$; 
deviations are parameterised in terms of the anomalous couplings $\Delta\kappa=\kappa-1$ and $\lambda$. 
The dependence of the total $W$-production cross section on these anomalous couplings is calculated using {\sc EPVEC}.
 For example, the upper limit of $0.58$~${\rm pb}$ for
hadronic $P_{T}>20$~${\rm GeV}$ corresponds
to the following $95\%$~C.L. limits on $\Delta\kappa$ and $\lambda$ :
\begin{eqnarray}
-4.7  \:\:<\:\:&\Delta\kappa&\:\:<\:\: 1.5 \:\:\:\: (\lambda=0) \:\:,\nonumber \\
-3.2  \:\:<\:\:&\lambda&\:\:<\:\: 3.2 \:\:\:\: (\Delta\kappa=0) \:\:. \nonumber
\end{eqnarray}
These limits on anomalous $WW\gamma$ couplings are insensitive to the assumed couplings at the $WWZ$ vertex due to the suppression by the propagator mass.
They are however significantly larger than the limits derived from analyses at the 
Tevatron~\cite{wgamma_d0} and LEP2~\cite{L3_limits}.

\section{Isolated-Track Search}

The H1 collaboration has recently reported the observation of six events containing an isolated high-$P_{T}$ 
lepton and large missing~$P_{T}$ in $36.5$~${\rm pb^{-1}}$ of $e^{+}p$ data~\cite{h1_paper}. The H1 search 
required isolated tracks with $P_{T}>10$~${\rm GeV}$ and a calorimeter $P_{T}$ exceeding $25$~${\rm GeV}$.
Five of the events contain muons, which may be compared to the $0.5$~$W$ events and $0.25$~other events 
(mainly $\gamma\gamma\to\mu^{+}\mu^{-}$) expected. 

\begin{figure}[!tb]
 \begin{center}
  \mbox{\epsfig{figure=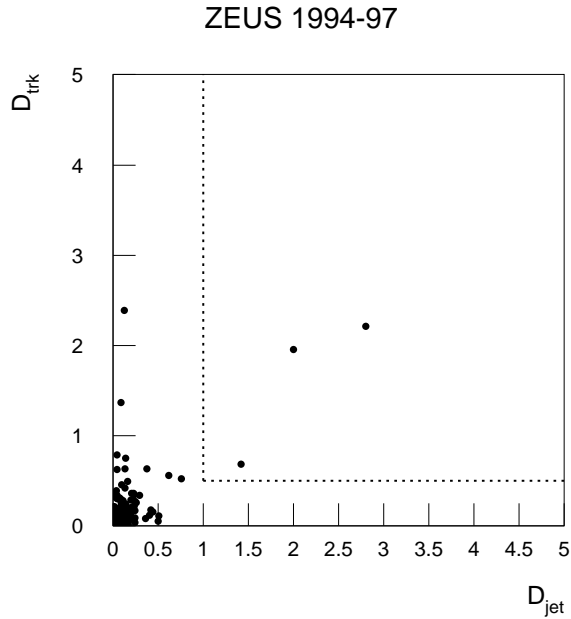}}
 \end{center}
\caption{{\em The jet and track $\{\eta,\phi\}$ isolation of high-$P_{T}$ tracks in events with large 
missing~$P_{T}$. The cuts used to select the tracks are described in the text. The dotted lines define the
region of interest.}}
\label{fig:iso_trk}
\end{figure}

Although the ZEUS data
presented above are in good agreement with Standard Model expectations, a separate search has been performed for
isolated high-$P_{T}$ tracks in events with a large missing~$P_{T}$, applying cuts similar to those 
used by H1. Because of the typical $10\%$ difference between observed and corrected hadronic-$P_{T}$ values, 
all events that have an uncorrected calorimeter $P_{T}$ greater than $20$~${\rm GeV}$ are selected. In addition, the
events are required to contain at least one jet with $E_{T}>5$~${\rm GeV}$, an electromagnetic fraction 
less than $0.9$ and an angular size greater than $0.1$~${\rm rad}$. Events with a neutral-current topology that have
an acoplanarity angle less than $0.2$~${\rm rad}$ are excluded.
The high-$P_{T}$ track must pass through at least 3 radial superlayers of the CTD (corresponding to 
$\theta\gtrsim  0.3$~${\rm rad}$) and have $\theta < 2.0$~${\rm rad}$.
The isolation variables ${\rm D_{jet}}$ and ${\rm D_{trk}}$ are defined for a given track as the $\{\eta,\phi\}$ 
separation of that track from the nearest jet and the nearest neighbouring track in the 
event, respectively. All tracks with $P_{T}>10$~${\rm GeV}$ in the selected events are plotted in the 
$\{\rm D_{trk},D_{jet}\}$ plane in Fig.~\ref{fig:iso_trk}. The $3$ tracks selected with ${\rm D_{trk}}>0.5$ and 
${\rm D_{jet}}>1.0$ agree well with the expectation of $5.7\pm 0.8$ tracks from combined Monte Carlo sources.
All three isolated tracks are positively charged and are identified as electrons using the algorithm and criteria
described above. This is consistent with the $3.5\pm 0.7$ ($2.0\pm 0.4$) electron type (muon type) events 
expected from Monte Carlo, of which $0.9$ ($0.4$) are from $W$ production.
Two of the isolated tracks correspond to events 2 and 3 of Table~\ref{tab:w_e_properties}. The third track is found 
in an event with neutral-current topology in which there is evidence of a large energy leakage into the backing 
calorimeter.
There is therefore no evidence of an excess of isolated high-$P_{T}$ tracks, whether identified as leptons or not,
in the $1994-1997$ ZEUS data.

\section{Conclusions}

A search for the decay $W\to e\nu$ in $e^{+}p$ collisions at a centre-of-mass energy of $300$~${\rm GeV}$ yields three candidate
events, of which $1.1\pm 0.3$ are estimated to arise from sources other than $W$ production. This results in an estimate of the
cross section for the process $e^{+}p\to e^{+}W^{\pm}X$ of $0.9^{+1.0}_{-0.7} \pm 0.2$~${\rm pb}$, consistent with the Standard Model 
prediction. The corresponding $95\%$~C.L. upper limit on the cross section of $3.3$~${\rm pb}$ is insensitive to 
uncertainties in the underlying hadronic-$P_{T}$ distribution. A search for the decay $W\to\mu\nu$ yields no candidate event, also
consistent with Standard Model expectations.
Events with large hadronic~$P_{T}$ in the combined electron plus muon sample
have been used to set $95\%$~C.L. upper limits on the cross section for $W$ production,
for example $0.58$~${\rm pb}$ for hadronic~$P_{T}$ greater than $20$~${\rm GeV}$.

A number of events with large missing $P_{T}$ and an isolated high-$P_{T}$ lepton, in excess of Standard Model expectations, has been 
reported by the H1 collaboration. The search presented in this paper, with similar cuts and sensitivity, has revealed no such excess.

\section*{Acknowledgements}

This work would not have been possible without the dedicated efforts of the HERA machine group and the DESY 
computing staff. We would also like to thank the DESY directorate for their strong support and encouragement throughout.
The design, construction and installation of the ZEUS detector would not have been possible without the hard work 
of many people who are not listed as authors. In addition, it is a pleasure to thank U.~Baur, D.~Zeppenfeld and M.~Spira for providing 
their calculations and for many fruitful discussions.

\end{document}